\documentclass[aps,pra,twocolumn]{revtex4}

\pdfoutput=1

\usepackage{amsmath}
\usepackage{amssymb}
\usepackage{amsbsy}
\usepackage{makeidx}
\usepackage{epsf}
\usepackage{graphicx}
\usepackage{amsfonts}
\usepackage{subfigure}
\usepackage{color}
\usepackage{epstopdf}

\newcommand{\pt}{$\mathcal{PT}$}

\begin{document}
\title{On the integrability of \pt-symmetric dimers}
\author{J.\ Pickton}
\author{H.\ Susanto}
\email{hadi.susanto@nottingham.ac.uk}
\affiliation{School of Mathematical Sciences, University of Nottingham, University Park, Nottingham, NG7 2RD, UK}


\begin{abstract}
The coupled discrete linear and Kerr nonlinear Schr\"odinger equations with gain and loss describing transport on dimers with parity-time (\pt) symmetric potentials are considered. The model is relevant among others to experiments in optical couplers and proposals on Bose-Einstein condensates in \pt symmetric double-well potentials. It is known that the models are integrable. Here, the integrability is exploited further to construct the phase-portraits of the system. A pendulum equation with a linear potential and a constant force for the phase-difference between the fields is obtained, which explains the presence of unbounded solutions above a critical threshold parameter. The behaviour of all solutions of the system, including changes in the topological structure of the phase-plane, is then discussed. 
\end{abstract}

\maketitle

\section{Introduction}

It is postulated in quantum physics that quantities we observe are eigenvalues of operators representing the dynamics of the quantities. Therefore, the energy spectra, i.e.\ the eigenvalues, are required to be real and bounded from below so that the system has a stable lowest-energy state. To satisfy such requirements, it was conjectured that the operators must be Hermitian. Non-Hermitian Hamiltonians have been commonly associated with complex eigenvalues and therefore decay of the quantities.

However, it turned out that hermiticity is not necessarily required by a Hamiltonian system to satisfy the postulate \cite{mois11}. Of particular examples have been systems exhibiting the so-called parity-time (\pt) symmetry, suggested by Bender and co-workers \cite{bend98,bend99,bend07}. A necessary condition for a Hamiltonian to be \pt  symmetric is that its potential $V(x)$ should satisfy the condition $V(x)=V^*(-x)$.

Optical analogues of such systems were proposed in \cite{rusc05,gana07,klai08} using two coupled waveguides with gain and loss. Note that such couplers were already studied in \cite{chen92,jorg93,jorg94}. The following successful experiments \cite{guo09,rute10} have stimulated extensive studies on \pt symmetric dimers, which are a finite-dimensional reduction of Schr\"odinger equations modelling, e.g., Bose-Einstein condensates with \pt symmetric double-well potentials \cite{li11,duan13,rame10,sukh10,miro11,wunn12,grae12,heis13,rodr12,dast13}. Nontrivial characteristics of the systems allow them to be exploited, e.g., for all-optical switching in the nonlinear regime, lowering the switching power and attaining sharper switching transition \cite{chen92} as well as a unidirectional optical valve \cite{rame10}. \pt symmetric analogues in coupled oscillators have also been proposed theoretically and experimentally recently \cite{schi11,rame12,lin12,schi12}. Note that coupled oscillators with gain and loss have already been considered in \cite{holm82}.

In this paper, we consider the following equations of motion \cite{li11,rame10,sukh10}
\begin{align}
\begin{split}
i\dot{u_1}=-u_2 -\delta |u_1|^2u_1 -i\gamma u_1,\\
i\dot{u_2}=-u_1 -\delta |u_2|^2u_2 +i\gamma u_2,
\end{split}
\label{gov}
\end{align}
where the dot represents differentiation with respect to the evolution variable, which is the propagation direction $z$ for nonlinear optics or the physical time $t$ for Bose-Einstein condensates, $\delta$ is the nonlinearity coefficient and $\gamma>0$ is the gain-loss parameter. Here, we consider two cases, i.e.\ when $\delta\neq0$ and $\delta=0$. For the former case, one can scale the coefficient such that $\delta=1$. It was shown numerically in \cite{sukh10} that the nonlinearity suppresses periodic time reversals of optical power exchanges between the sites, leading to the symmetry breaking and a sharp beam switching to the waveguide with gain.

When $\gamma=0$, Eq.\ \eqref{gov} has two conserved quantities
\begin{align}
P&=|u_1|^2+|u_2|^2,\label{P}\\
H&=-\frac\delta2\left(|u_1|^4+|u_2|^4\right)-P,
\end{align}
which are commonly referred to as the power and the Hamiltonian/energy, respectively. Using the Liouville-Arnold theorem (or Liouville-Mineur-Arnold theorem) \cite{arno88,libe87}, \eqref{gov} is integrable since the degree of freedom is equal to the number of conserved quantities. By defining the site-occupation probability difference
\begin{equation}
\Delta=|u_2|^2-|u_1|^2, \label{D}
\end{equation}
Kenkre and Campbell \cite{kenk86} showed that $\Delta$ satisfies a $\phi^4$-equation, which explains the presence of Josephson tunneling and self-trapped states, with the latter corresponding to $\Delta$ being sign-definite, as well as the transition between them. It was later shown \cite{jens85,cruz88} that $\Delta$ also satisfies the pendulum equation.

When $\gamma\neq0$, Eq.\ \eqref{gov} is still integrable \cite{rame10}. System \eqref{gov} is actually a special case of a notably integrable dimer derived in \cite{jorg93,{jorg94},{jorg98}} (see also a brief review of integrable oligomers in \cite{susa09}). The conserved quantities in that case are
\begin{align}
r&=u_1u_2^*+u_1^*u_2,\\
H&=-\frac\delta2\left(|u_1|^4+|u_2|^4\right)-P-i\gamma\left(u_1u_2^*-u_1^*u_2\right).
\end{align}
It was reported that the general system could be reduced to a first-order differential equation with polynomial nonlinearity and it possesses blow-up solutions that was observed numerically  \cite{jorg93,{jorg94}}.

In this paper, we consider \eqref{gov} and show that it can be reduced to a pendulum equation with a linear potential and a constant drive. The same equation has been obtained recently, parallel to and independently from this work, by Kevrekidis, Pelinovsky and Tyugin \cite{kevr13} and by Barashenkov, Jackson and Flach \cite{bara13} through a different formalism. The linear potential and constant drive explain the presence of unbounded solutions. We exploit the strong relation between the problem and the geometry of circles.  We also discuss the qualitative pictures of all solutions of the governing equations.

In Section \ref{sec2}, we rewrite the governing equations \eqref{gov} in terms of power, population imbalance, and phase difference between the wavefields in the channels. In the section, we also derive a constant of motion. In Section \ref{sec3}, we analyse the characteristics of the fixed points, which are the time-independent solutions of the system. In Section \ref{sec4}, we reduce our system in Section \ref{sec2} further into one equation. Here, we show that the problem is described by a pendulum equation with a linear potential and a constant drive. In the section, we analyse the pendulum equation qualitatively through its phase-portrait. In Section \ref{sec5}, we discuss the phase-portrait of the system that is composed of trajectories with the same value of a constant of motion. The constant corresponds to power, that is a conserved quantity when $\gamma=0$. The case of linear systems is discussed in Section \ref{sec6}. Finally, we conclude our work in Section \ref{conc}.

\section{Governing equations in polar forms}
\label{sec2}

Writing $u_1$ and $u_2$ in polar form
\begin{eqnarray}
u_j=|u_j|e^{i\phi _j},\,j=1,2,\label{polar}
\end{eqnarray}
and defining the variable of phase difference between $u_1$ and $u_2$
\begin{equation}
\theta=\phi _2 - \phi_1,\label{t}
\end{equation}
the equations of motion \eqref{gov} can then be expressed in terms of \eqref{P}, \eqref{D} and \eqref{t} as (see the Appendix)
\begin{align}
\dot{P}&=2\gamma\Delta,\label{eq:Pdot}\\
\dot{\Delta}&=2\gamma P + 2\sqrt{P^2-\Delta^2}\sin\theta,
\label{eq:Deltadot}\\
\dot{\theta}&=\Delta\left(1-\frac{2\cos\theta}{\sqrt{P^2-\Delta^2}}\right).
\label{eq:thetadot}
\end{align}
We can limit the phase difference to be in the interval $-\pi\leq\theta<\pi$. Note that the argument angle $\phi_j$ (and hence $\theta$) is undefined when $|u_j|=0$ (or $|u_1||u_2|=0$).

Taking $\gamma\neq 0$, the conditions for equilibrium points are
\begin{eqnarray}
\Delta=0 ,\quad \sin\theta+\gamma=0. 
\label{eq:eqpt}
\end{eqnarray}
This shows that no equilibrium points can exist with $\Delta\neq 0$, therefore demonstrating the non-existence of the self trapped state that is observed for the case when $\gamma=0$. When $\gamma>1$ it follows that no equilibrium points exist. This is a threshold of total \pt-symmetry breaking where no periodic solutions can exist. However, unbounded trajectories will be shown to always exist for any $\gamma>0$.

\begin{figure}[tbhp]
(a)\includegraphics[width=7.0cm,clip=]{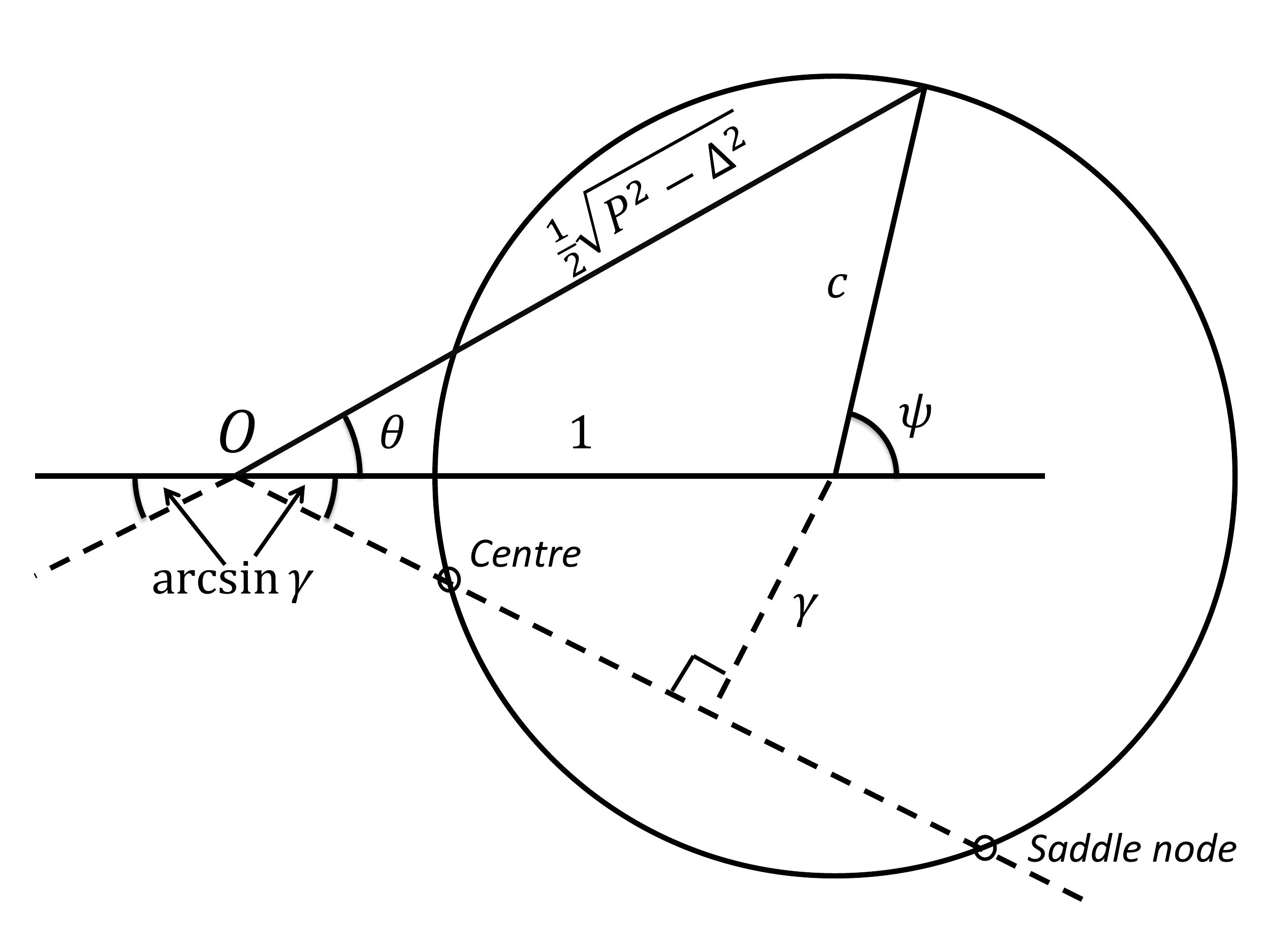}   
(b)\includegraphics[width=7.0cm,clip=]{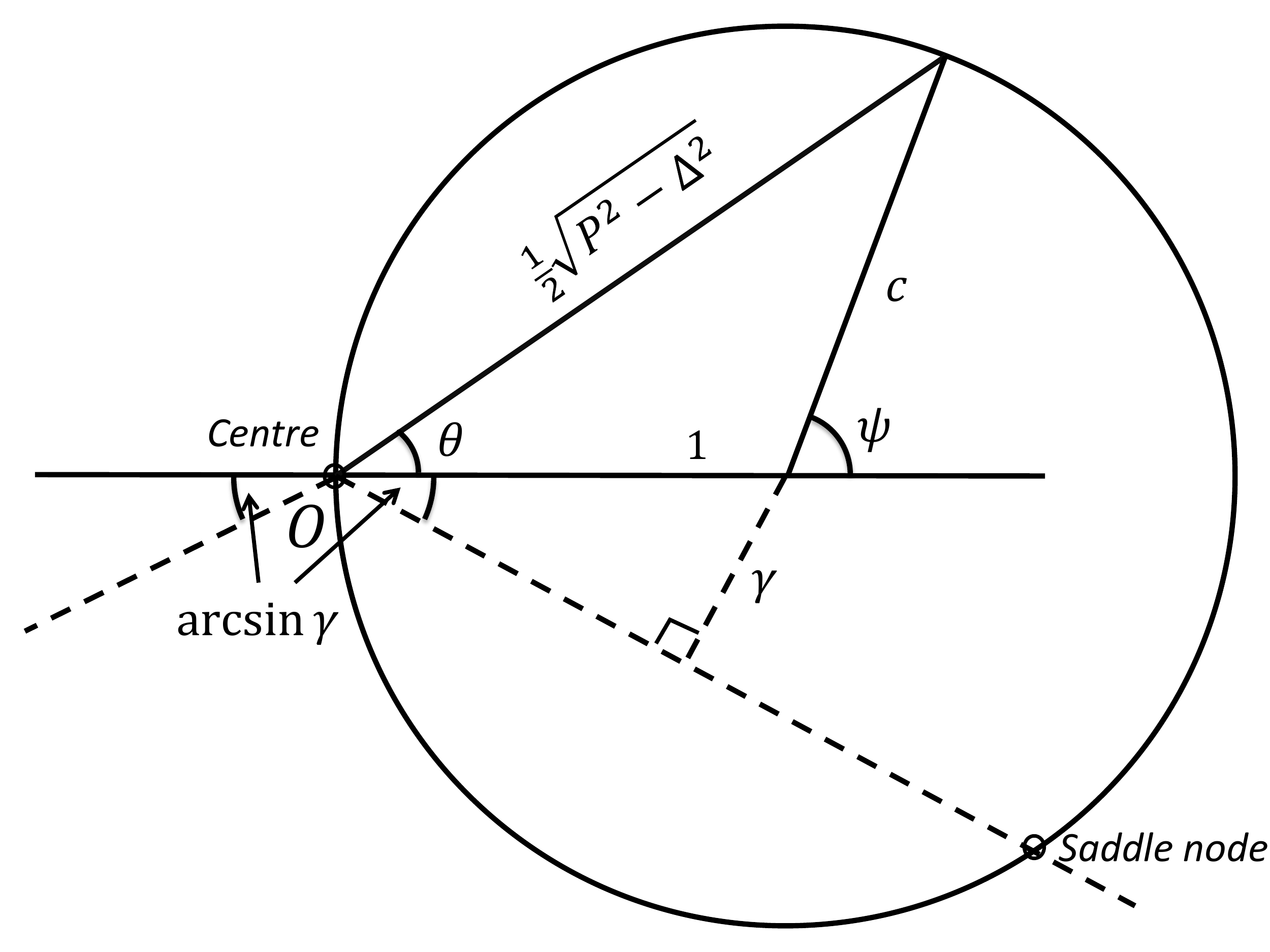}   
(c)\includegraphics[width=7.0cm,clip=]{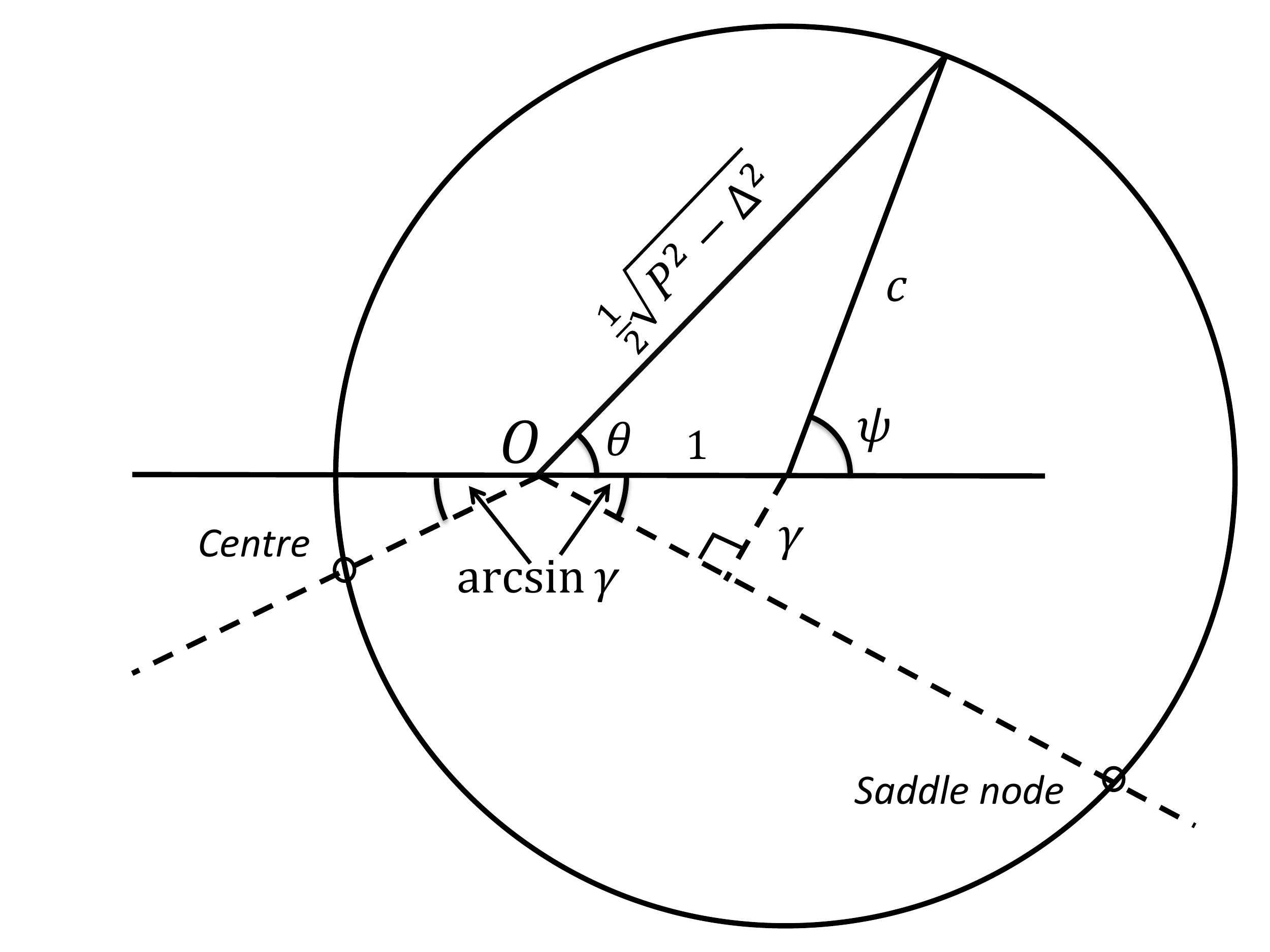}   
\caption{Geometrical interpretation of \eqref{eq:c} (cf.\ \eqref{ad1}) as a triangle. Sketches of the circles of radius $c$ are also shown for the case of (a) $\gamma<c<1$, (b) $\gamma<c=1$, and (c) $\gamma<1<c$, obtained from \eqref{eq:c} by varying $\theta$. The dashed lines show the angles, such that intersections with the circles would indicate the positions of the equilibrium points.}
\label{sk2}
\end{figure}

After some manipulations (see the Appendix for the details), it is possible to find a constant of motion of \eqref{eq:Pdot}, \eqref{eq:Deltadot}, and \eqref{eq:thetadot} that is given by
\begin{eqnarray}
\frac{1}{2}\sqrt{P^2-\Delta^2}\left( \frac{1}{2}\sqrt{P^2-\Delta^2}-2\cos\theta \right)=c^2-1,
\label{eq:c}
\end{eqnarray}
where $c$ is a constant. We can impose the constant to be non-negative with no loss of generality. It is important to note that one can draw similarities of \eqref{eq:c} to the cosine rule for triangles with three edge length: $1$, $c$ and $\frac{1}{2}\sqrt{P^2-\Delta^2}$ (see Fig\ \ref{sk2}).


By varying $\theta$, trajectories of \eqref{eq:c} form circles with radius $c$ centred on a point a unit distance from the origin. Three different sketches corresponding to different values of $c$ are given in Fig.\ \ref{sk2}.

Equation (\ref{eq:c}) can also be rearranged to give
\begin{eqnarray}
\sqrt{P^2-\Delta^2}=2\cos\theta \pm 2\sqrt{c^2-\sin ^2 \theta},\label{conmo}
\end{eqnarray}
from which a few things can be said about the constant $c$ and how this affects what values $\theta$ can take. Firstly $c$ must be real to give a real solution. If $c=0$ then $\theta =0$. If $0<c<1$ then $\theta$ is bounded such that $-\arcsin c < \theta < \arcsin c$. The '$-$' solution in \eqref{conmo} and \eqref{peq} can exist only for $c$ within this range. {Only when $c=1$, $P^2$ can equal $\Delta ^2$}, in which case $-\pi /2<\theta <\pi/2$. If $c>1$ then $\theta$ is unbounded.


\section{The equilibrium points}
\label{sec3}

It is always natural to first analyse the behaviour of the time-independent solutions. Using the conditions for the equilibrium points Eq.\ (\ref{eq:eqpt}) alongside (\ref{conmo}), we can find the value of the power, $P$, at equilibrium points for different values of $c$, 
i.e.\
\begin{align}
P_l&=2\sqrt{1-\gamma^2} \pm 2\sqrt{c^2-\gamma^2}. 
\label{peq}
\end{align}
Applying the inequality $P\geq\sqrt{P^2-\Delta^2}$ to Eq.\ \eqref{conmo}, Eq.\ \eqref{peq} is also the minimum power that a (generally time-dependent) solution can attain.

The value $c=0$ is particular as explicit solutions can be found. In this case $|u_1|=Ae^{-\gamma t}$, $|u_2|=\frac{1}{A}e^{\gamma t}$ and $\theta=0$, where $A$ is a constant determined by the initial conditions. This gives one example of where \pt-symmetry is broken for any non-zero $\gamma$.

If $c\in (0,\gamma )$, no equilibrium points can exist as the power \eqref{peq} is complex-valued. 
Hence, all solutions are unbounded.

When $c=\gamma$, an equilibrium point lies on the boundary of $\theta$ with $\theta = -\arcsin\gamma$. 

For trajectories with $c\in (\gamma ,1)$, there are two equilibrium points at $\theta = -\arcsin\gamma$. This is shown in Fig.\ \ref{sk2} as the intersections of the circles and the dashed lines given by Eq.\ (\ref{eq:eqpt}). The first of these is a centre 
and the second is a saddle node with the corresponding power $P_{centre}$ and $P_{saddle}$, respectively. 
Exploiting the geometry of the triangles and circles in Fig.\ \ref{sk2}, one can obtain that 
\begin{eqnarray}
\frac{1}{2}P_{centre} = \sqrt{1-\gamma^2}-\sqrt{c^2-\gamma^2},
\label{eq:Pcentre} \\
\frac{1}{2}P_{saddle} = \sqrt{1-\gamma^2}+\sqrt{c^2-\gamma^2}.
\label{eq:Psaddle}
\end{eqnarray}

For the case of $c=1$, the trivial equilibrium point at $P_{centre}=0$ is found to represent a centre for these trajectories. There is also an equilibrium point located at $\theta=-\arcsin\gamma$ and $P_{saddle}=4\sqrt{1-\gamma^2}$, that can be straightforwardly shown to be a saddle node.

Solutions with $c>1$ also have two equilibrium points located at $\theta=-\arcsin\gamma$ and $\theta=-\pi+\arcsin\gamma$. The former represents a saddle node with $P_{saddle}=2\sqrt{1-\gamma^2}+2\sqrt{c^2-\gamma^2}$ and the latter is a centre with $P_{centre}=2\sqrt{c^2-\gamma^2}-2\sqrt{1-\gamma^2}$.

From the results above, we obtain that generally speaking the larger $c$ is, the larger the power of stable solutions can be.

\section{Further reduction of the equations of motion}
\label{sec4}

\begin{figure}[tbhp]
(a)\includegraphics[width=7.0cm,clip=]{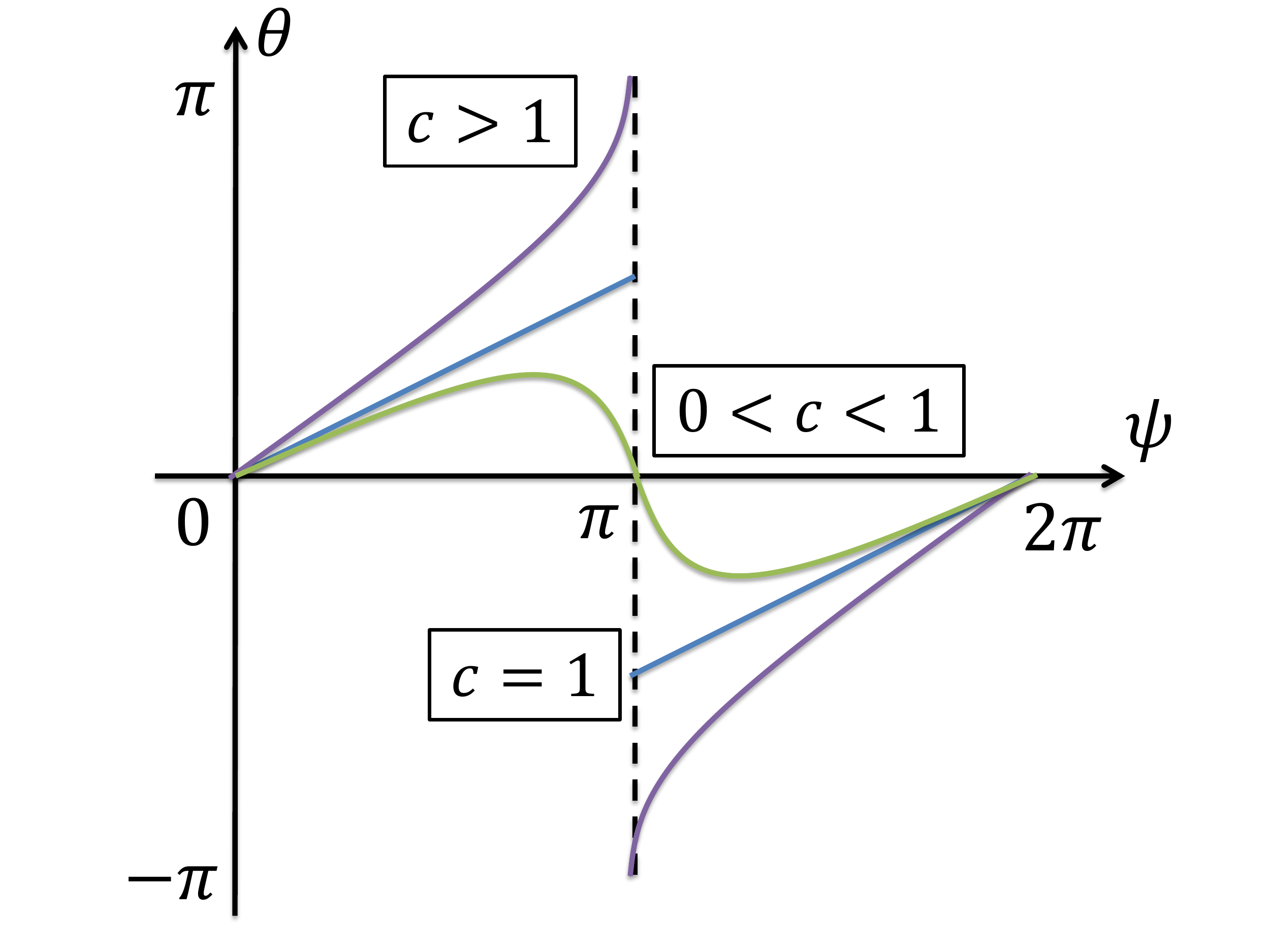}   
(b)\includegraphics[width=7.0cm,clip=]{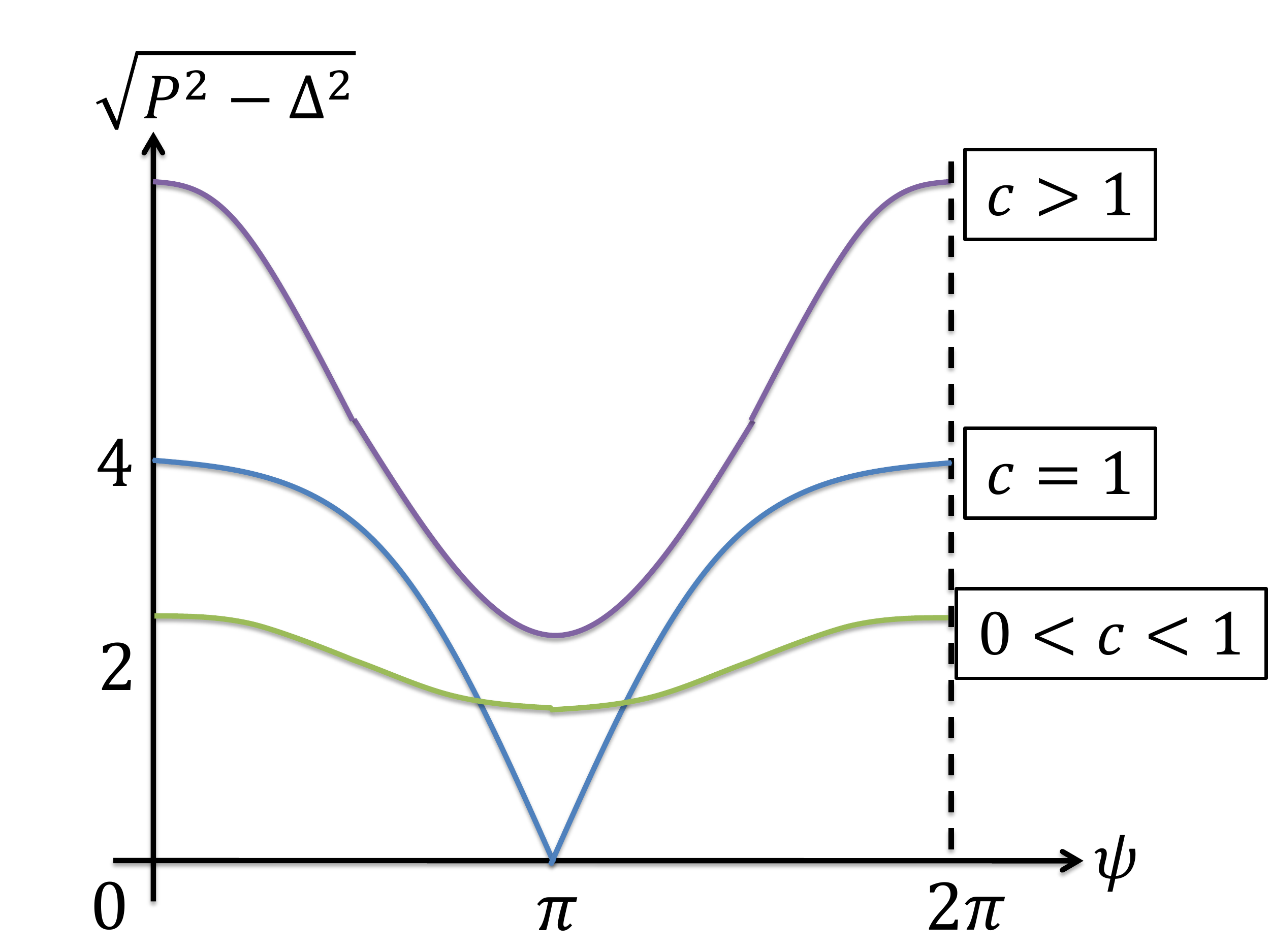}   
\caption{(Colour on-line) Sketches of (a) $\theta$ against $\psi$, (b) $\sqrt{P^2-\Delta ^2}$ against $\psi$ for three different values of $c$.}
\label{sk4}
\end{figure}

In order to analyse the general solutions of the governing equations \eqref{gov}, it is useful to define a new variable $\psi$, i.e.\ the arc angle, that parametrises the circles as indicated in Fig.\ \ref{sk2}. The variable can be expressed by
\begin{eqnarray}
c\sin\psi = \frac{1}{2}\sqrt{P^2-\Delta^2}\sin\theta,
\label{eq:cpsi1} \\
c\cos\psi = \frac{1}{2}\sqrt{P^2-\Delta^2}\cos\theta -1,
\label{eq:cpsi2}
\end{eqnarray}
where $0\leq c$ and $0\leq \psi< 2\pi$. These equations can be rearranged to yield 
\begin{eqnarray}
\sqrt{P^2-\Delta^2}=2\sqrt{1+c^2+2c\cos\psi},
\label{eq:cpsi3} \\
\tan\theta = \frac{c\sin\psi}{c\cos\psi + 1}.
\label{eq:cpsi4}
\end{eqnarray}
The relationships are plotted in Fig.\ \ref{sk4}. Using the remark following \eqref{peq}, note that \eqref{eq:cpsi3} is a lower bound to the power of solutions. After carrying out some rearranging and differentiation (see the Appendix), we find that
\begin{eqnarray}
{\Delta = \dot{\psi}} ,
\label{eq:d} \\
P=2\gamma\psi + 2k ,
\label{eq:k}
\end{eqnarray}
where $k$ is, technically, a constant of motion. We use the word "technically" because care must be taken in how $\psi$ is defined. Because we are choosing to define $0\leq\psi< 2\pi$ we must remember that as a trajectory crosses this boundary, the value of $k$ changes by an amount $2\pi\gamma$ to ensure that the power, $P$, remains continuous. If $\psi$ should be defined unbounded then $k$ would indeed remain constant for all time. It so happens that for $\gamma\neq 0$, $k$ has similarities to $P$ with $\gamma=0$, where it is well known that the power remains constant for all trajectories.

The true value of defining the variable $\psi$ and the constant $c$ is made apparent when we substitute {$\dot{\psi}=\Delta$} with equations (\ref{eq:cpsi1}) and (\ref{eq:k}) into equation (\ref{eq:Deltadot}). The equations of motion then reduce to a second order differential equation
\begin{eqnarray}
\frac{1}{4}\ddot{\psi}=\gamma\left(\gamma\psi+k\right)+c\sin\psi .
\label{eq:oscillator1}
\end{eqnarray}
This is a main result of this paper. When $\gamma =0$, we indeed obtain the equation for a pendulum similarly to \cite{jens85,cruz88} that was derived through a different approach. The presence of $\gamma\neq0$ introduces a linear potential and a constant drive into the pendulum equation. It may be possible to solve Eq.\ (\ref{eq:oscillator1}) in terms of Jacobi elliptic functions. Asymptotic solutions of the equation were derived in \cite{kevr13}. However, instead we will study the qualitative behaviours of the solutions in the ($\psi$,$\dot{\psi}$)-phase plane.


\section{General solutions}
\label{secb}

\begin{figure}[tbhp]
(a)\includegraphics[width=7.0cm,clip=]{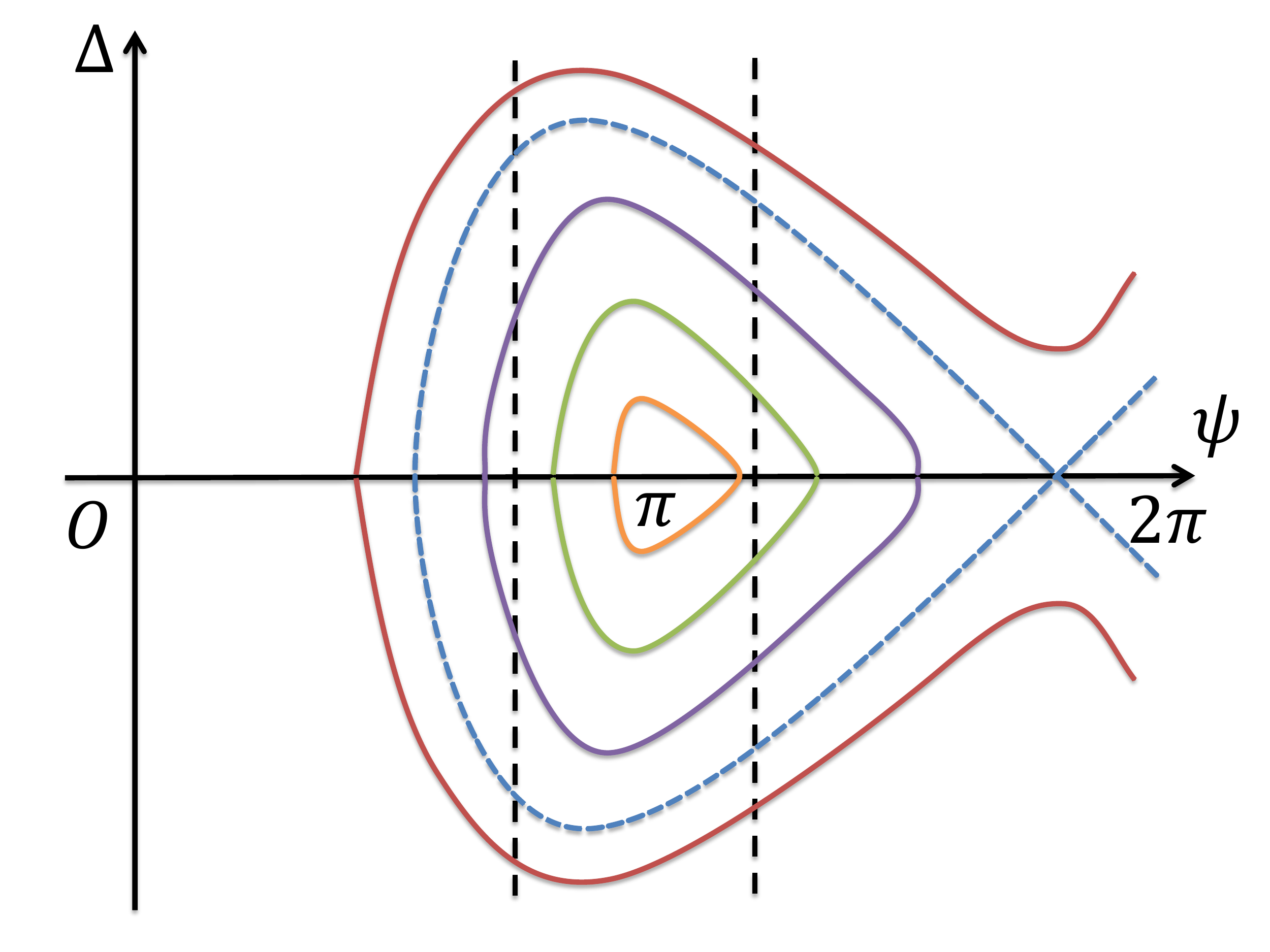}   
(b)\includegraphics[width=7.0cm,clip=]{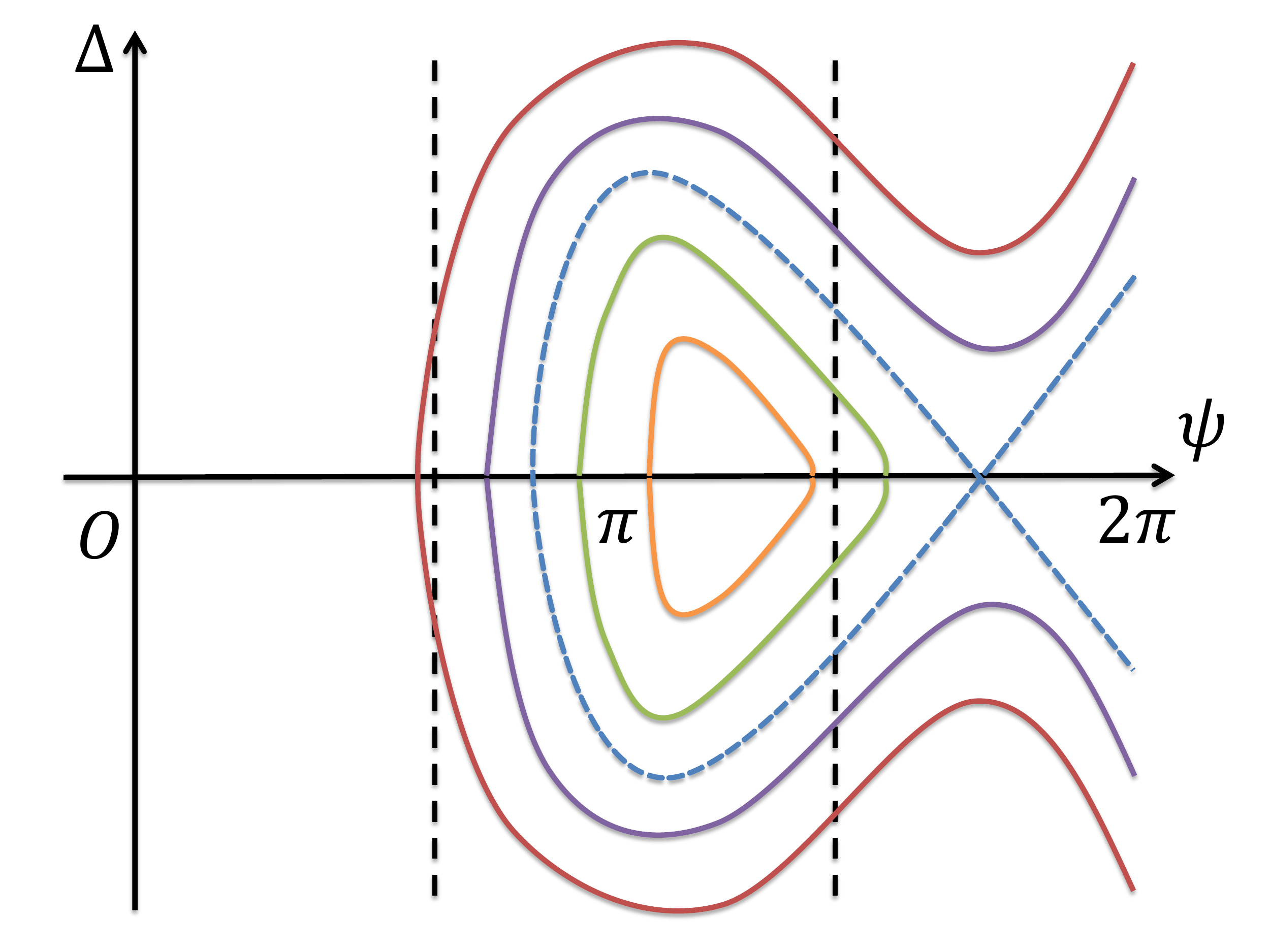}   
\caption{{(Colour on-line)} Sketches of the trajectories in the $(\psi,\dot{\psi}=\Delta)$-phase plane for $c\in (0,1)$. See the text for the details. 
}
\label{sk5}
\end{figure}

The first integral of Eq.\ (\ref{eq:oscillator1}) can be obtained by using Eq.\ (\ref{eq:cpsi3}), i.e.\
\begin{eqnarray}
\frac{1}{4}\dot{\psi}^2 = \left(\gamma\psi+k\right)^2 -2c\cos\psi -c^2-1.
\label{eq:oscillator2}
\end{eqnarray}
The phase-portrait can then be plotted rather easily and is demonstrated in Fig.~\ref{sk5} for $c\in(\gamma,1)$. The phase plane can have two topologically different structures, {one with and one without a stable region}.

\begin{figure}[tbhp]
(a)\includegraphics[width=8.0cm,clip=]{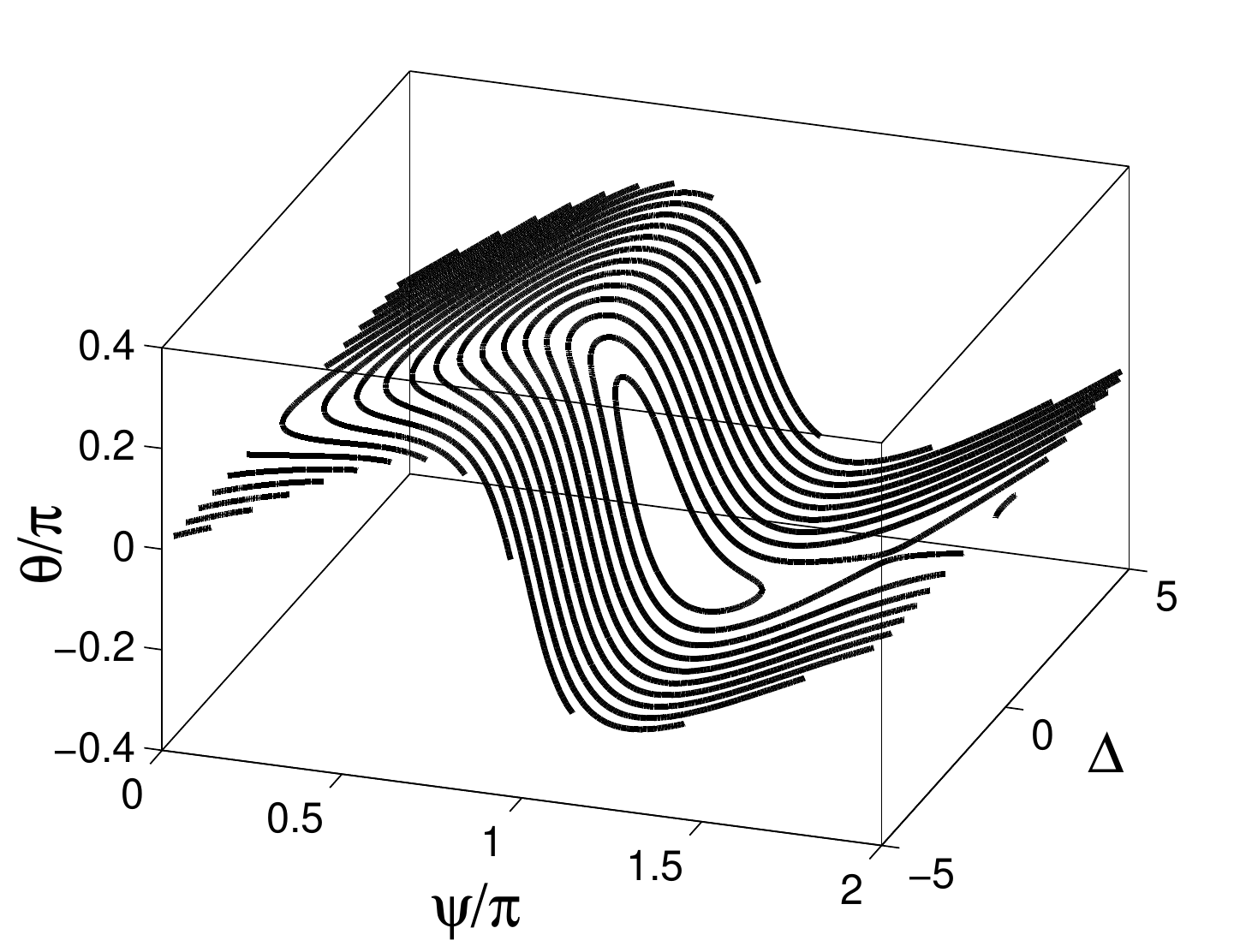}   
(b)\includegraphics[width=8.0cm,clip=]{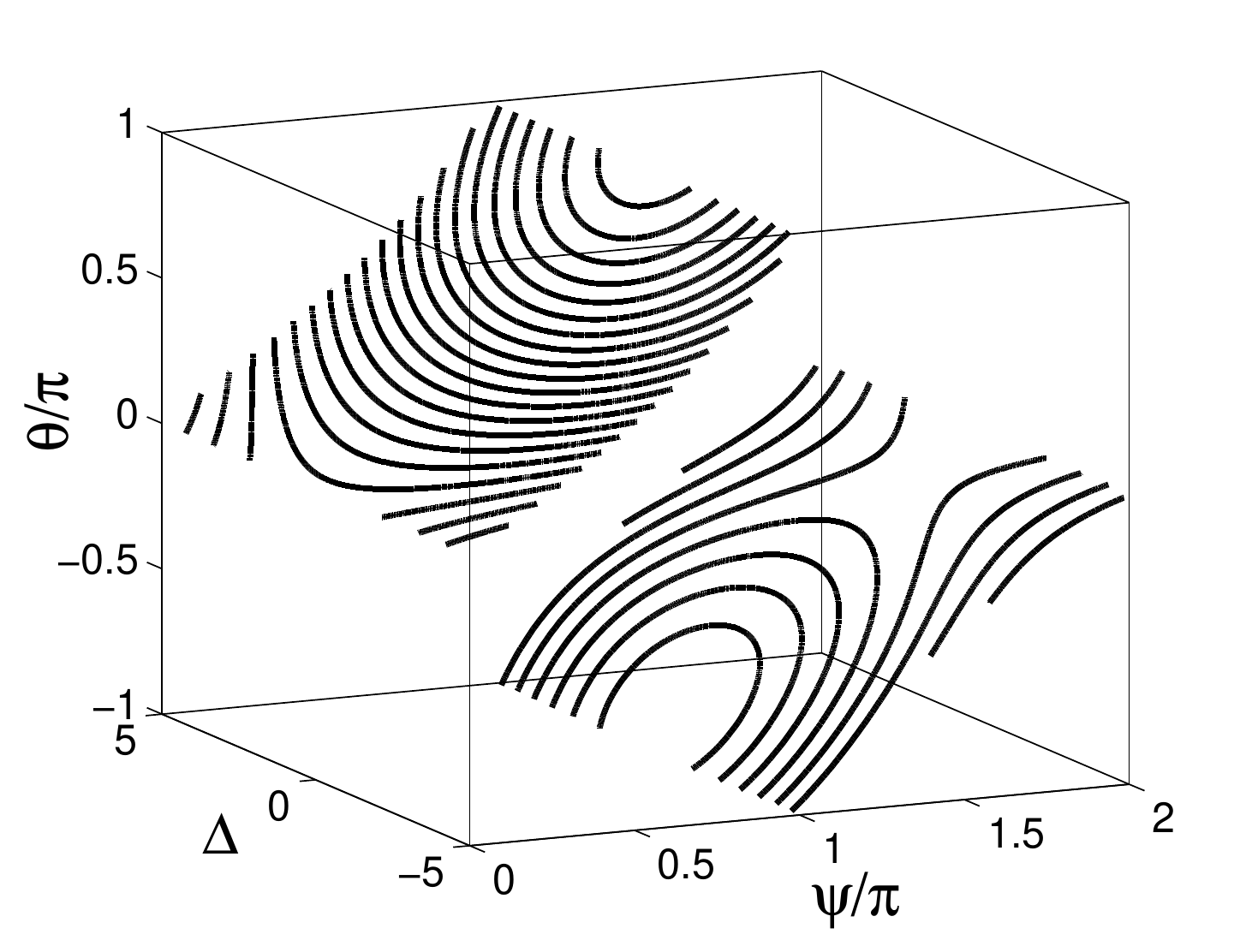}
\caption{{(Colour on-line)} Plots of trajectories in the ($\psi$,$\Delta,\theta$)-phase space for $\gamma=0.4$ and (a) $c=0.7$, (b) $c=3$.}
\label{padd1}
\end{figure}

{Fig.\ \ref{padd1} shows how this phase plane, combined with Fig.~\ref{sk4}a (cf.\ \eqref{eq:cpsi4}) can be plotted in a three-dimensional graph. This then gives an idea of how solutions appear in the $(\theta ,\Delta)$ phase plane.}

{When $c\in(0,1)$ there are boundaries on $\theta$ which correspond to $\psi = \pi\pm\arccos c$. These are represented by vertical dashed lines in Fig.\ \ref{sk5}. Trajectories that cross these boundaries will behave differently, in the $(\theta,\Delta)$ plane, from those trajectories which do not. This results in a more diverse set of solutions when $c\in(0,1)$ than for otherwise. One can note that when $c=1$ these vertical dashed lines merge at $\psi=\pi$ to correspond to the point where $\theta$ is in fact undefined.}

{It is useful to know the value of the constant $k$ for the trajectories that touch the boundary at $\psi=\pi\pm\arccos c$. Such values are found by using Eq.\ (\ref{eq:k}) and Eq.\ (\ref{eq:cpsi3}) to be
\begin{eqnarray}
k_{\pm} = -\gamma\pi + \sqrt{1-c^2} \mp \gamma \arccos c .
\label{eq:kboundary}
\end{eqnarray}}

{How a trajectory behaves depends on how its $k$-value compares to the four main values of $k$, namely $k_{saddle}$, $k_{centre}$, $k_+$ and $k_-$, where
\begin{eqnarray}
k_{centre} = \frac{1}{2}P_{centre} - \gamma \psi_{centre},
\label{eq:kcentre} \\
k_{saddle} = \frac{1}{2}P_{saddle} - \gamma \psi_{saddle},
\label{eq:ksaddle}
\end{eqnarray}}
with $\psi_{centre}$ and $\psi_{saddle}$ being the values of $\psi$ at the equilibrium points, i.e.\
\begin{eqnarray}
\psi_{centre} = \pi + |\arccos\gamma - \arccos\frac{\gamma}{c}|,
\label{eq:Psicentre} \\
\psi_{saddle} = \pi + \arccos\gamma + \arccos\frac{\gamma}{c},
\label{eq:Psisaddle}
\end{eqnarray}
which can be obtained using \eqref{eq:cpsi3}-\eqref{eq:cpsi4} (or similarly exploiting the geometric pictures in Fig.\ \ref{sk2}). By plotting out these values of $k$ against $c$ for a fixed $\gamma$ one sees that there is a point where $k_{saddle}$ and $k_-$ cross. 
Three types of solution exist for all $c\in(0,1)$. The first of these is a stable trajectory that never touches a boundary for $\theta$. This corresponds to $k\in(k_{centre},k_+)$. Second are trajectories with $k\in(k_+,min(k_-,k_{saddle}))$. These cross one boundary of $\theta$. The final type is an unstable trajectory with $k>max(k_-,k_{saddle})$.

\begin{figure*}[bthp]
(a)\includegraphics[width=8.0cm,clip=]{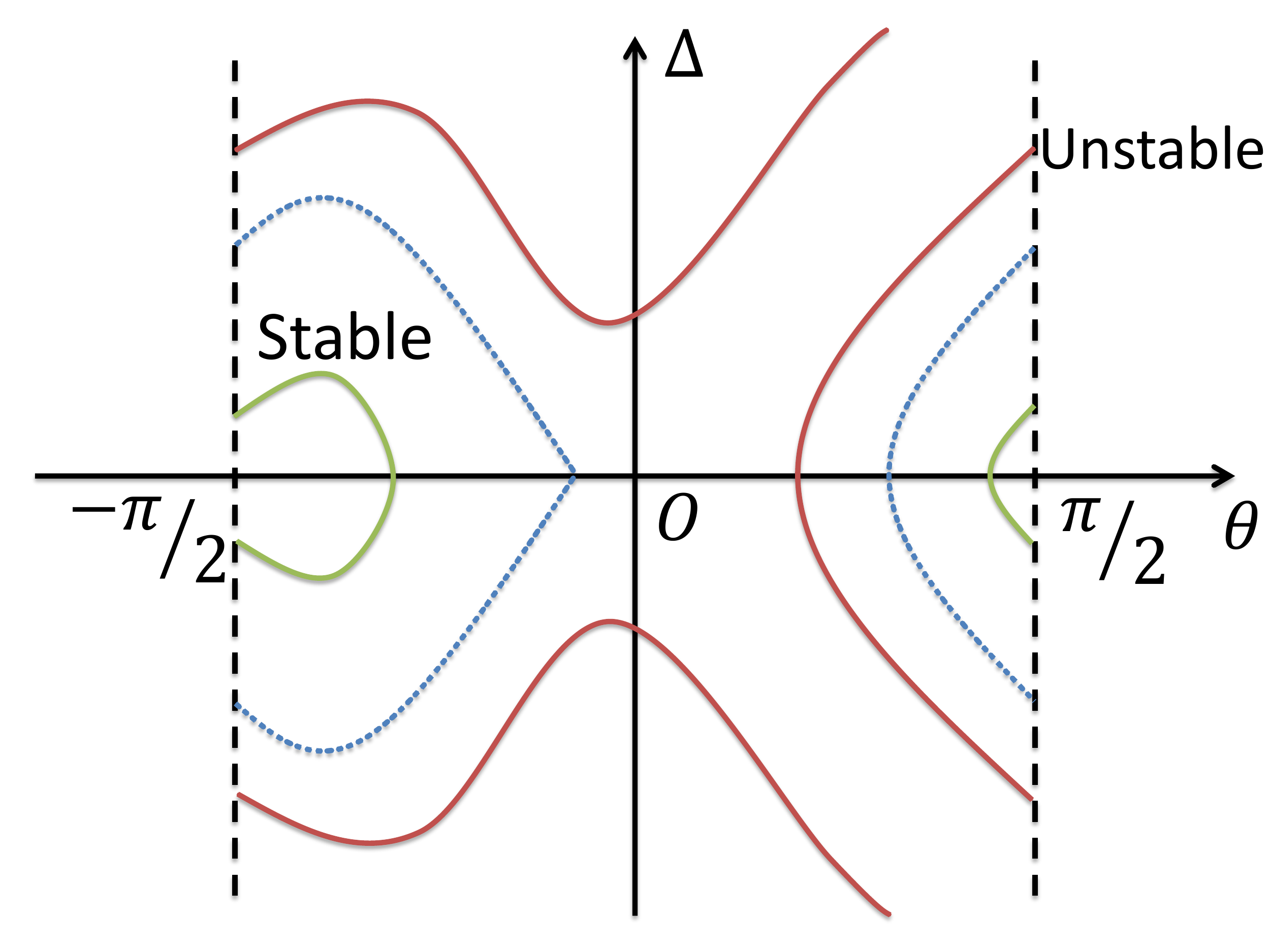}   
(b)\includegraphics[width=8.0cm,clip=]{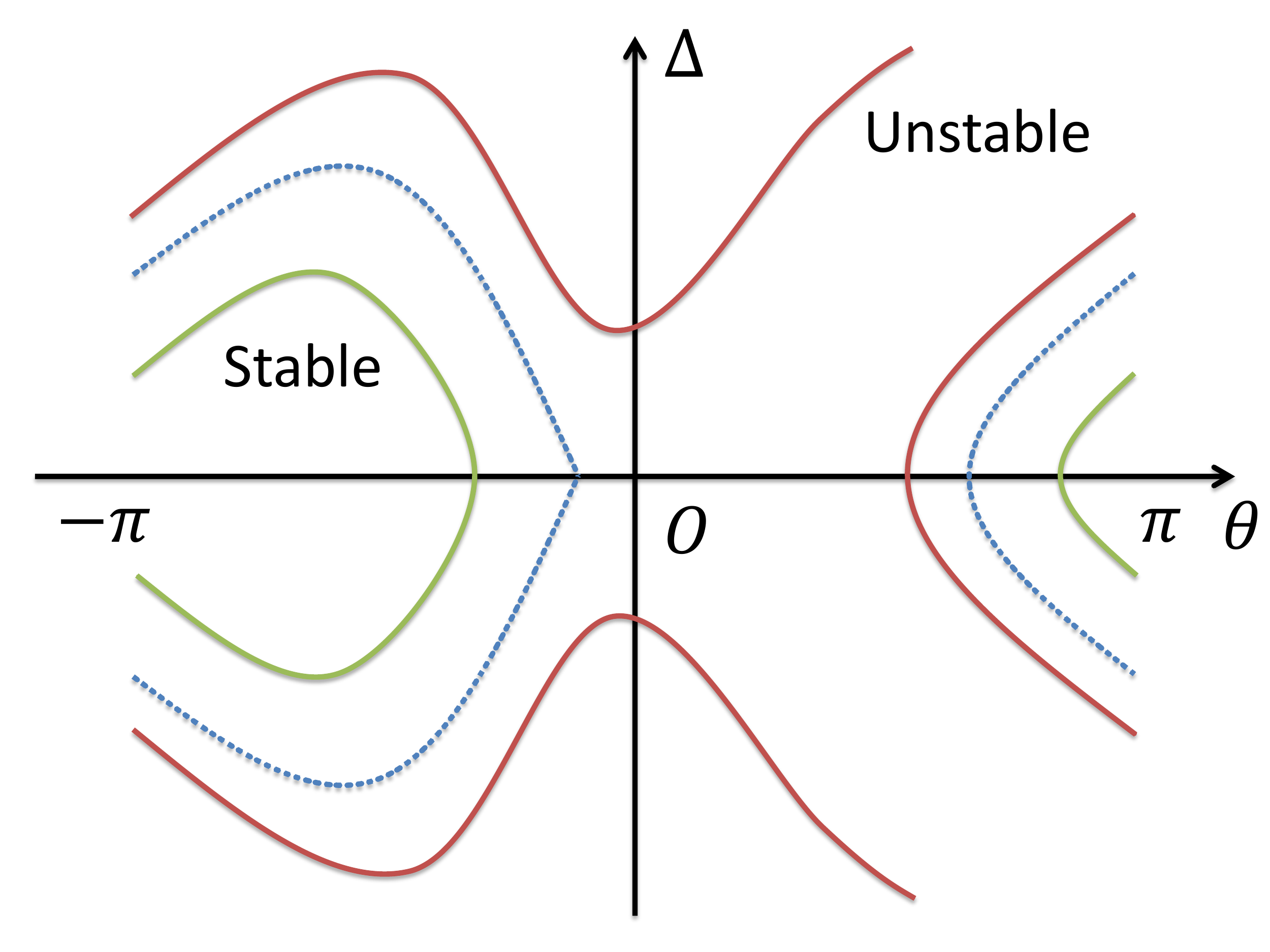}\\   
(c)\includegraphics[width=8.0cm,clip=]{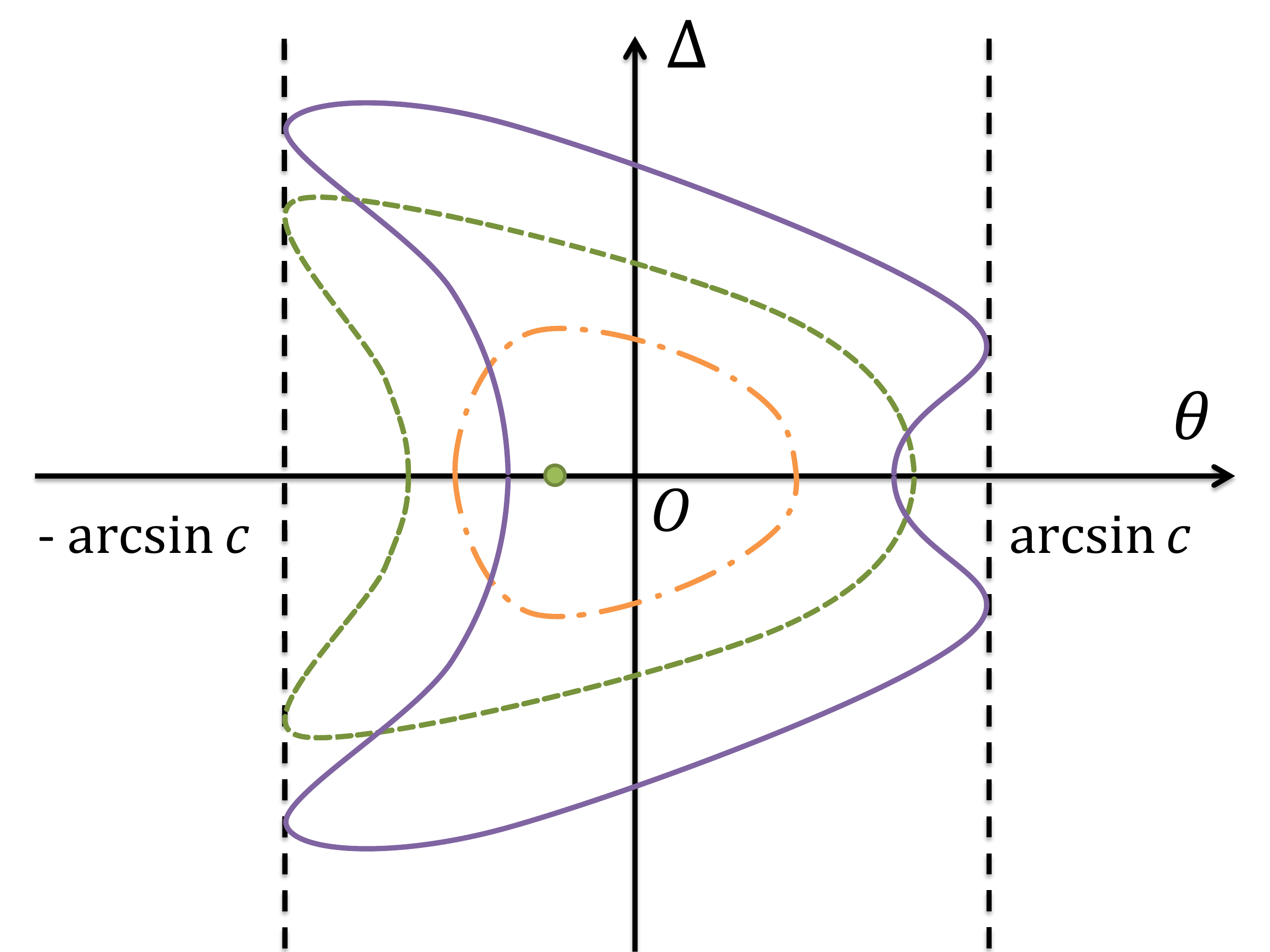}   
(d)\includegraphics[width=8.0cm,clip=]{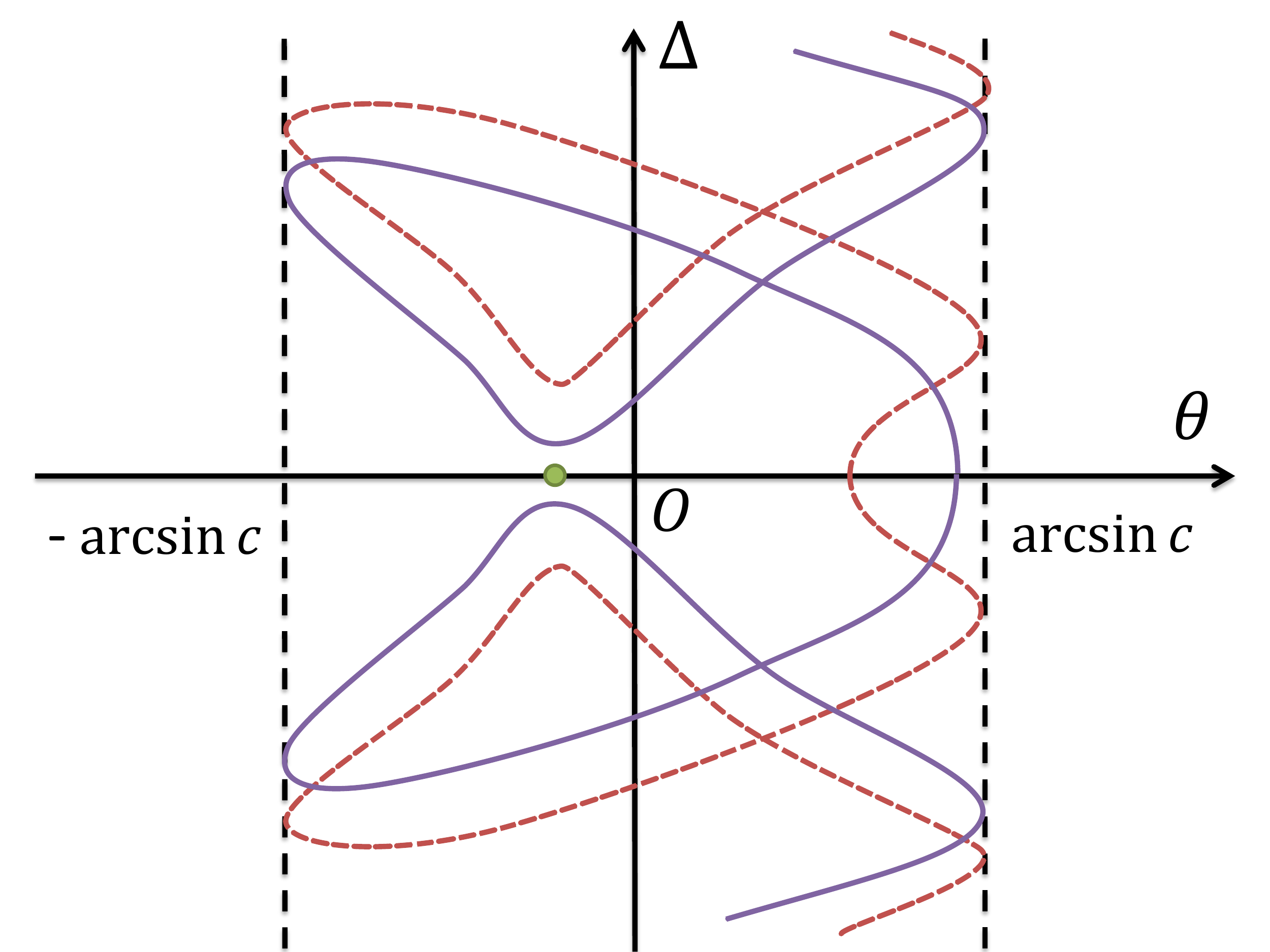}   
\caption{{(Colour on-line)} Sketches of trajectories in the ($\theta$,$\Delta$)-phase plane for (a) $c=1$ and (b) $c>1$. The case $c\in (\gamma ,1)$ has (c) periodic and (d) unbounded solutions that are sketched in different panels for clarity. The (purple) solid trajectories in (c) and (d) cannot co-exist for the same values of $c$, as explained in the text.}
\label{sk3}
\end{figure*}

{In addition to the trajectories above, there are two types of solutions that cannot coexist for the same $c$-value. This is because one solution exists when $k_- < k_{saddle}$ and the other when $k_- > k_{saddle}$, which is obvious that both conditions cannot be met at the same time. When $k_- < k_{saddle}$, there is additionally another type of stable trajectories that crosses both boundaries of $\theta$. This is demonstrated in Fig.~\ref{sk5}a. However, if $k_- > k_{saddle}$ then instead there is another type of unstable trajectory, as shown in in Fig.~\ref{sk5}b. All of the different types of trajectory are sketched in Fig.~\ref{sk3}.}

\section{Trajectories with the same 'power'}
\label{sec5}

In the conservative case $\gamma=0$, the power $P$ is independent of time, see \eqref{P}. In that case, plots in Figs.\ \ref{padd1} and \ref{sk3} correspond to varying power, i.e.\ each trajectory has different value of $P$, and a fixed 'energy', i.e.\ the same value of first integral \eqref{eq:oscillator2}. However, it is not the common practice as usually one plots trajectories corresponding to the same value of power.

In the general case when $\gamma\neq0$, $P$ is no longer constant. It is therefore not possible to compose a similar phase portrait consisting of trajectories with constant $P$. However, we find that one could instead plot phase-portraits with the same value of $k$, which interestingly runs parallel with the different solutions observed at the same values of $P$ for $\gamma=0$, see \eqref{eq:k}.

When comparing periodic trajectories that all have the same value of 'power' $k$, it is useful to study the ($\psi$,$P$)-phase plane, which qualitatively can be obtained by plotting and analysing \eqref{eq:cpsi3} (see Fig.\ \ref{sk4}b) and \eqref{eq:k}. By doing this, we find that different values of $k$ can give different results in the ($\theta$,$\Delta$)-phase plane.

There are three important $k$ values to consider. The first is the minimum value $k$ can take. The second is the point where the equilibrium point at $\theta=-\arcsin\gamma$ bifurcates from a centre to a saddle node. This is found from substituting $c=\gamma$ into Eq.\ (\ref{eq:ksaddle}). The third important value of $k$ is when the separatrix passing through the saddle point has a $c$-value of one, found from substituting $c=1$ into Eq.\ (\ref{eq:ksaddle}). These values of $k$ are given respectively by
\begin{align}
k_0&=-\pi\gamma ,\label{k1}\\
k_\gamma &=-\pi\gamma -\gamma\arccos\gamma +\sqrt{1-\gamma^2},\label{k2}\\
k_1&=-\pi\gamma -2\gamma\arccos\gamma +2\sqrt{1-\gamma^2}.\label{k3}
\end{align}
Substituting $\gamma=0$ into the above expressions gives the important values of $k$ as 0, 1 and 2. Realising that here $P=2k$ makes clear that these are generalisations of the critical values found for $P$ when $\gamma=0$, where $P=0,2,4$ corresponds respectively to the presence of only trivial solution, the emergence of a pair of fixed points $\Delta\neq0$, and the emergence of rotational (or running) states \cite{ragh99}. When $\gamma=1$ all three of these values are equal, which is the threshold of total \pt-symmetry breaking, i.e.\ there are no periodic, bounded solutions for $\gamma>1$. Note that for any $|\gamma|<1$, there are two equilibrium points. The $c$-values of the equilibria give the upper and lower bounds the periodic paths can have for a certain $k$.

\begin{figure}[tbhp]
(a)\includegraphics[width=7.0cm,clip=]{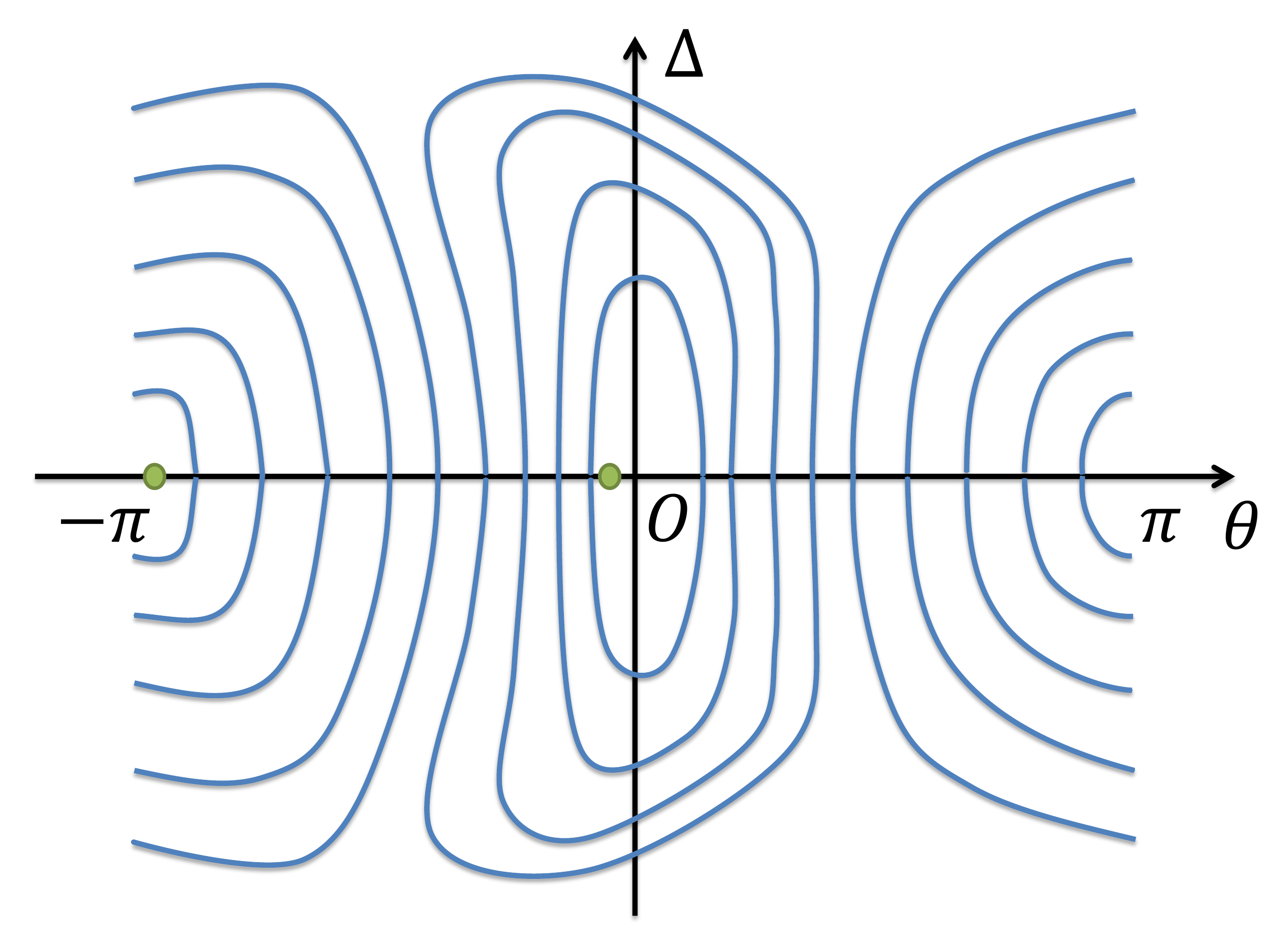}   
(b)\includegraphics[width=7.0cm,clip=]{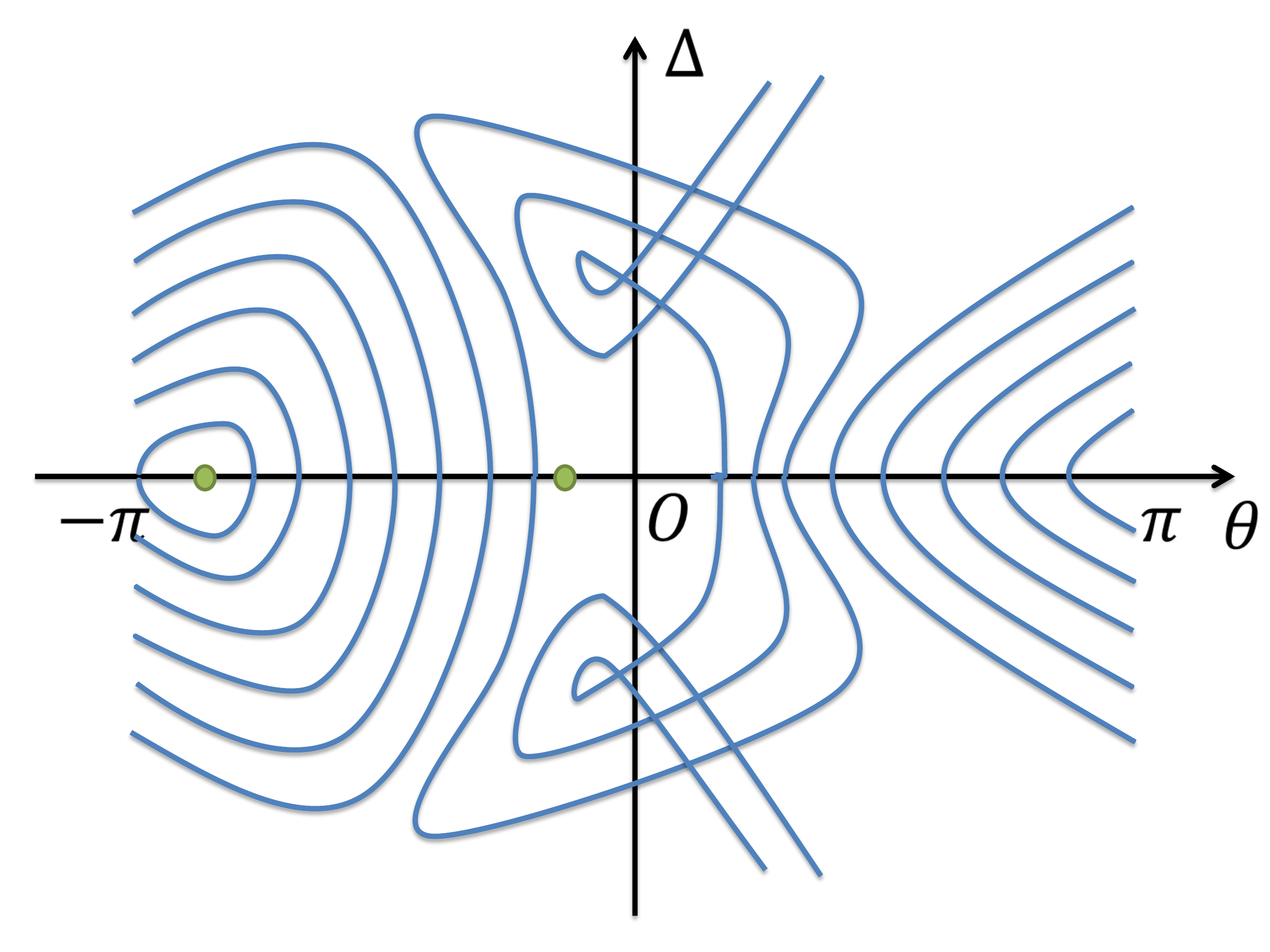}   
(c)\includegraphics[width=7.0cm,clip=]{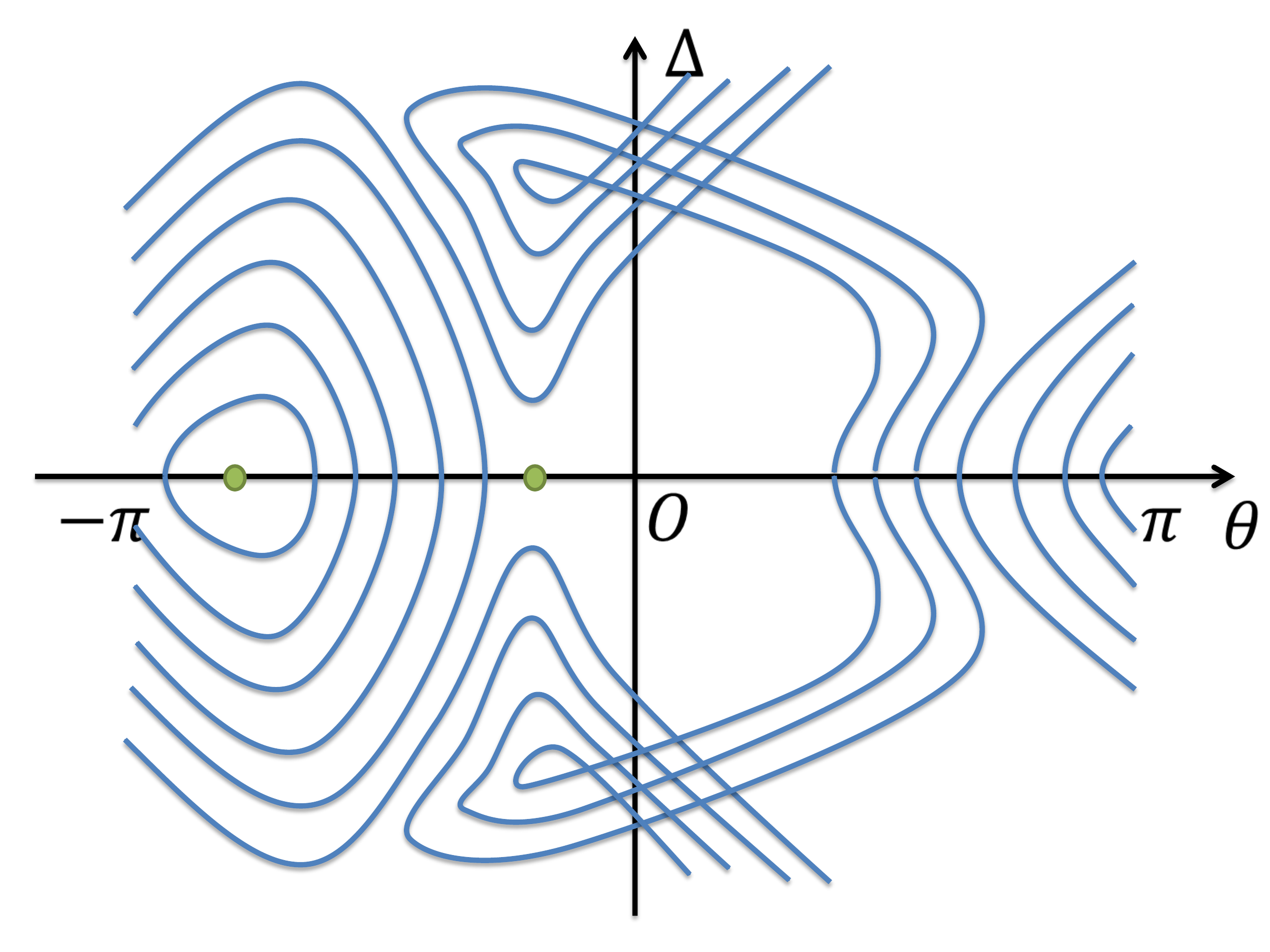}   
\caption{{(Colour on-line)} Sketches of the ($\theta$,$\Delta$)-phase plane when $k$ is constant. (a) All solutions are bounded, with $k\in (k_0,k_\gamma)$. (b) Unbounded solutions exist, with $k\in (k_\gamma ,k_1)$. (c) Unbounded solutions exist for some $c>1$, with $k>k_1$.}
\label{sk6}
\end{figure}

In Fig.\ \ref{sk6}, we sketch the possible phase-portraits of the system for $\gamma\neq0$ within three different intervals determined by the critical values of $k$ above. Note that the figures are actually composed of trajectories taken from the phase portraits with different values of $c$ sketched in Fig.\ \ref{sk3}. They can also be seen as the dynamical regimes for arbitrary initial conditions, showing the regions of periodic and unbounded solutions, that we refer to here as stability and instability regions. Similar figures obtained using comprehensive numerical computations were presented in \cite{sukh10}.

In panel (a), all the solutions are periodic. As $k$ increases, there will be a critical value \eqref{k2} above which the only equilibrium point becomes unstable. When $\gamma=0$, this is where a pair of fixed points with nonzero $\Delta$ emerges. For $\gamma\neq0$, instead we have unbounded solutions. The instability region is nevertheless contained within the region of stable, periodic solutions. When $k$ is increased further passing \eqref{k3}, the stability region that contains the instability region vanishes completely.

Sukhorukov, Xu, and Kivshar \cite{sukh10} also presented the dependence of the minimal input intensity on the gain/loss coefficient and the corresponding phase difference required for nonlinear switching, i.e.\ the minimum intensity above which the solutions would be unbounded. According to the analytical results presented herein, those would correspond to the separatrices of \eqref{eq:oscillator2} (see Fig.\ \ref{sk5}). Analytical expressions should be derivable, that we leave for the interested reader.

\section{Linear Equations}
\label{sec6}

After analysing the nonlinear dimer \eqref{gov} with $\delta\neq0$, finally we discuss the linear equations described by \eqref{gov} with $\delta=0$. This is particularly interesting because the system has been realised experimentally in \cite{guo09,rute10}.

Transforming the equations into the polar forms by similarly defining $P$, $\Delta$ and $\theta$, we obtain the same equations except (\ref{eq:thetadot}), which is now given by
\begin{eqnarray}
\dot{\theta}=\Delta\left(-\frac{2\cos\theta}{\sqrt{P^2-\Delta^2}}\right).
\end{eqnarray}

Following the similar reduction, we obtain that instead of \eqref{conmo} the constant of motion is given by
\begin{eqnarray}
\sqrt{P^2-\Delta ^2}\cos\theta = 2c.\label{add1}
\end{eqnarray}
This, instead of circles, is an equation for a straight line passing a distance $c$ from the origin. By parametrising the line with the new variable $\psi$ defined by
\begin{eqnarray}
\psi = -c\tan\theta,\label{add2}
\end{eqnarray}
we find that (cf.\ \eqref{psidot} and \eqref{eq:k})
\begin{eqnarray}
\dot{\psi}=\Delta,\,P = 2\gamma\psi + 2k, \label{add3}
\end{eqnarray}
where $k$ is also a constant of integration. Using the above equations combined with the identity $\sqrt{P^2-\Delta^2}\sin\theta = -\psi$ (obtained from \eqref{add1} and \eqref{add2}) in equation (\ref{eq:Deltadot}) gives
\begin{eqnarray}
\frac{1}{4}\ddot{\psi}+\left(1-\gamma ^2\right)\psi = k\gamma .
\end{eqnarray}
This is a forced, simple harmonic equation and therefore can be solved explicitly to give $\psi$ in terms of $t$. 
Due to the second equation in \eqref{add3}, we therefore can conclude that the oscillating power reported in, e.g., \cite{gana07,klai08} has internal frequency $\omega=2\sqrt{1-\gamma^2}$.

\section{Conclusion}
\label{conc}

We have studied analytically linear and nonlinear \pt-symmetric dimers, where we described the whole dynamics of the system. The effect of nonlinearity that induces \pt-symmetry breaking for gain/loss parameters below that of the linear system, which in a previous work was referred to as nonlinear suppression of time-reversals \cite{sukh10}, has been analysed as well. Our analytical study may offer a new insight into the global dynamics of directional waveguide couplers with balanced gain and loss or Bose-Einstein condensates in a double-well potential with a balanced sink and source of atoms. In additions to our qualitative analysis, one could extend the study here to the analytical expression of the solutions of the nonlinear \pt-symmetric dimer \eqref{eq:oscillator2}, that may be expressed in terms of Jacobi elliptic functions. In that case, the oscillation frequency of the power in the nonlinear system would be obtained. Here, we only derive the frequency of the linear system.

\appendix

\section{Detailed calculations}

Differentiating the expression \eqref{polar} gives
\begin{eqnarray}
\dot{u}_j=\left(\frac{d}{dt}|u_j|+i|u_j|\frac{d\phi _j}{dt}\right)e^{i\phi _j},
\end{eqnarray}
which by substituting into the governing equations \eqref{gov} yields
\begin{eqnarray}
\frac{d}{dt}|u_1|=&-|u_2|\sin\theta -\gamma |u_1|, \\
\frac{d}{dt}|u_2|=&|u_1|\sin\theta +\gamma |u_2|,
\end{eqnarray}
from the real parts and
\begin{eqnarray}
|u_1|\frac{d\phi _1}{dt} =&|u_2|\cos\theta +|u_1|^3, \\
|u_2|\frac{d\phi _2}{dt} =&|u_1|\cos\theta +|u_2|^3,
\end{eqnarray}
from the imaginary parts.

Using the equations above, we can then write
\begin{eqnarray}
\frac{d\theta}{dt}=\left(|u_1|^2-|u_2|^2\right)\left(\frac{\cos\theta}{|u_1||u_2|}-1\right).
\end{eqnarray}
Using the product rule for differentiation we can also obtain the following four equations
\begin{align}
\frac{d}{dt}\left(|u_1|^2\right)&=-2|u_1||u_2|\sin\theta -2\gamma |u_1|^2, \\
\frac{d}{dt}\left(|u_2|^2\right)&=2|u_1||u_2|\sin\theta +2\gamma |u_2|^2, \\
\frac{d}{dt}\left(|u_1||u_2|\right)&=\left(|u_1|^2-|u_2|^2\right)\sin\theta,\label{ad2} \\
\frac{d}{dt}\left(|u_1||u_2|\cos\theta\right)& =|u_1||u_2|\left(|u_1|^2-|u_2|^2\right)\sin\theta\label{ad3}.
\end{align}
Using the above equations and noting that
\begin{equation}
\sqrt{P^2-\Delta ^2}=2|u_1||u_2|,
\end{equation}
Eqs.\ \eqref{eq:Pdot},\eqref{eq:Deltadot}, and \eqref{eq:thetadot} can be immediately obtained.

Next, one can compare \eqref{ad2} and \eqref{ad3} to obtain that
\begin{eqnarray}
\frac{d}{dt}\left(|u_1|^2|u_2|^2\right)=2\frac{d}{dt}\left(|u_1||u_2|\cos\theta\right) .
\end{eqnarray}
Therefore, we can define a constant $c$ such that
\begin{eqnarray}
c^2=1+|u_1|^2|u_2|^2-2|u_1||u_2|\cos\theta,\label{ad1}
\end{eqnarray}
which is nothing else but Eq.\ \eqref{eq:c}. 

After parametrisation of the constant of motion by $\psi$, i.e.\ Eqs.\ \eqref{eq:cpsi1}-\eqref{eq:cpsi2}, one can differentiate \eqref{eq:cpsi2} to obtain that
\begin{eqnarray}
-c\dot{\psi}\sin\psi  = |u_1||u_2|\left(|u_1|^2-|u_2|^2\right)\sin\theta .
\end{eqnarray}
Now combining it with \eqref{eq:cpsi1} yields
\begin{eqnarray}
\dot{\psi}=|u_2|^2-|u_1|^2=\Delta.
\label{psidot}
\end{eqnarray}
Using \eqref{eq:Pdot}, we see that
\begin{eqnarray}
\dot{P}=2\gamma\dot{\psi},
\end{eqnarray}
which can be integrated to give \eqref{eq:k}. 

\end{document}